\documentclass[12pt,epsf]{article}
\def\be{\begin{equation}}
\def\ee{\end{equation}}
\def\bea{\begin{eqnarray}}
\def\eea{\end{eqnarray}}

\def\be{\begin{equation}}
\def\ee{\end{equation}}
\def\bea{\begin{eqnarray}}
\def\eea{\end{eqnarray}}
\def\half{\frac{1}{2}}

\def\case#1/#2{\textstyle\frac{#1}{#2}}
\def\k0{\kappa_{0}}

\begin{document}
\begin{titlepage}

\vspace{.7in}

\begin{center}
\Large
{\bf Penrose Limits, the Colliding Plane Wave Problem and the Classical String Backgrounds.}\\
\vspace{.7in}
\normalsize
\large{ 
 Alexander Feinstein
}\\
\normalsize
\vspace{.4in}

{\em Dpto. de F\'{\i}sica Te\'orica, Universidad del Pa\'{\i}s Vasco, \\
Apdo. 644, E-48080, Bilbao, Spain}\\
\vspace{.2in}
\end{center}
\vspace{.3in}
\baselineskip=24pt

\begin{abstract}
\noindent 
We show how the Szekeres form of the line element is naturally adapted to study
Penrose limits in classical string backgrounds.  Relating the ``old" colliding
wave problem to the Penrose limiting procedure as employed in string theory we
discuss how two orthogonal Penrose limits uniquely determine the underlying target space
when certain symmetry is imposed.
We construct a conformally deformed  background with two distinct, yet
exactly solvable in terms of the string theory on R-R backgrounds, Penrose limits.
Exploiting further the similarities between the two problems we find  that the Penrose
limit of the gauged WZW Nappi-Witten universe is itself a gauged WZW plane wave solution of Sfetsos 
and Tseytlin. Finally, we discuss some issues related to singularity, show the
existence of a large class of non-Hausdorff solutions with Killing Cauchy Horizons and indicate a
 possible
resolution of the problem of the definition of quantum vacuum  in string theory on these time-dependent backgrounds.  
\end{abstract}

\vspace{.3in}

\end{titlepage}

Penrose limit \cite{penrose}, \cite{guven}, a limiting procedure applied to a null geodesic in a 
generic spacetime and resulting in a p-p wave, has attracted considerable attention
recently \cite{blau, bf2, gursoy, kobi, hubney} . The main interest stems from the fact that string theory is exactly
solvable for certain plane waves both in NS-NS and R-R backgrounds \cite{papers2}, \cite{papers3}.
 Another point of interest is the relation, via the Penrose limit,  of the maximally supersymmetric
backgrounds of the M-theory, the so-called $ AdS \times S$  solutions with those of the $IIB$ superstring
theory \cite{bf2}. These relations have inspired an interesting derivation of the spectrum of 
$IIB$ theory on Minkowski and maximally supersymmetric Hpp-wave spacetimes using the 
$AdS/CFT$ correspondence \cite{berenstein} for the holographic dual of $d=4$, $N=4$ Super-Yang-Mills theory.

The main purpose of this Letter is to establish and  exploit the relation between the well-studied problem of the
collision of plane waves in General Relativity and the Penrose limiting procedure, as presented
in the context of the string theory.

To keep the discussion in a way  familiar to relativists we will stick to $4$-dimensional
space times. The generalisation to higher dimensions is quite straight forward
 and will not add qualitatively to the issues discussed in this paper. Our starting point
will be the so-called $AdS_{2}\times S^{2}$ spacetime, called otherwise the Bertotti-Robinson
solution \cite{bertotti}, \cite{robinson}:
\be
ds^2={q^2\over r^2}(-dt^2+dr^2+r^2d\theta^2+r^2\sin^2\theta
d\phi^2) \label{b-r}
\ee

The line element (\ref{b-r}) is a non-singular solution of Einstein-Maxwell equations with a uniform
electromagnetic field such that $F_{ab}F^{ab}=q^2=constant$. In  higher dimensions
 the electromagnetic filed is generalised to p-form potentials.
In the context of supergravity theory, this solution was discussed in \cite{k-g}.

The Bertotti-Robinson universe has been also a subject of an extensive study in the context
of the so-called colliding plane wave problem in General Relativity \cite{griffiths} (and the references
therein).
 The idea is that the
line element (\ref{b-r}) can be seen as produced in a non-linear interaction (collision) of two plane
electromagnetic  waves with constant, not necessarily equal, profiles. 

To put the metric into the form convenient from the point of view of the colliding wave
problem one performs the following co-ordinate transformation ( see for example \cite{griffiths})

\be
\begin{array}{rcl}
t + r & = & \coth [{1\over 2}\;{\rm sech}^{-1}(\cos(au+bv))-
{y\over 2q}]\\*[8pt]
t - r & = & -\tanh [{1\over 2}\;{\rm sech}^{-
1}(\cos(au+bv))+{y\over 2q}]\\*[8pt]
\theta & = & \pi/2 -(bv-au) \label{conftrans}\\*[8pt]
\phi & = & x/q,
\end{array}
\ee

\noindent where $q={1\over\sqrt{2ab}}$. With this we obtain:

\be
ds^{2}=-2du dv+\cos^{2}[au-bv] dx^{2}+
\cos^{2}[au+bv] dy^{2} \label{b-s}
\ee

The  positive
constants $a$ and $b$ are related to the strengths of the incoming electromagnetic
plane waves. In these co-ordinates, and extended into the plane wave regions by introducing
$u \to u\Theta(u)$ and  $v \to v\Theta(v)$, where $\Theta(u)$ and
$\Theta(v)$  are the usual Heaviside functions, the line element is called the Bell-Szekeres
solution \cite{b-s}.
For more details on this solution in the context of the  classical colliding wave problem 
 the reader is addressed to \cite{b-s, m-t, c-h} and the monograph  \cite{griffiths}.
The quantum field theory on the Bell-Szekeres background
was discussed in \cite {fein-ps}. 

Now, to obtain the Penrose limit for the Bell-Szekeres 
(Bertotti-Robinson- $AdS_{2}\times S^{2}$)
 spacetime, 
in the sense used in string  theory, becomes quite simple. Put  the parameters $a=0$,
or $b=0$ and we get plane wave limiting behaviour, both in the metric and the electromagnetic
 field. Note, that these Penrose limits formally co-incide with the incoming plane waves
 producing the interaction region (3). In the colliding wave problem, the parameters $a$ and $b$ are related to
the focal lengths of the plane waves $L_{1}\sim  a^{-1}$ and $L_{2}\sim b^{-1}$, which in turn
measure the strength of the incoming waves. Long focal length corresponds to 
a weak wave and vice versa. 
It is easy
 to show that the procedure to obtain the limit is equivalent to the scaling introduced in the string theory 
context  \cite{guven, blau, bf2}. We will not expand on this issue here, but will just mention
that in practice, 
to obtain the limiting behaviour of the metric one may also consider dropping the dependence
of the metric on $u$ or on $v$ \cite{bf2}, or, which is equivalent, 
substituting $u=0$ or $v=0$, or in turn equivalent to introducing the Heavyside functions as in
colliding wave approach.
We will therefore
refer to these limits the v-limit ($u=0$) and the u-limit ($v=0$) respectively. It is this observation
which takes us to explore the relation between the colliding wave problem and the notion of the 
Penrose limit further. 

In general, the Penrose limit may be evaluated for an arbitrary spacetime. In practice, however, it
is sufficient, especially  with the string theory in the back of our mind, 
to concentrate on spacetimes with at least two commuting spacelike Killing
vectors ($N-2$ Killing vectors in a general higher dimensional case). To simplify
 our discussion  further we will be concentrating on those spacetimes for which the Killing vectors
are mutually orthogonal. This will keep the $4$-dimensional line element diagonal, and will
allow us to cast the metric into the following so-called Szekeres form \cite{szekeres}:

\be
ds^2=- f(u,v)\,dudv + G(u,v)\,(e^{p(u,v)}\,dx^2 +e^{-p(u,v)}\,dy^2).\label{cw}
\ee

Any 4-dimensional spacetime with two orthogonal Killing directions can be put into the
above form,  spacetime (\ref{b-s})  being a special case.
 For pure gravitational fields, or in the case of massless fields such as dilaton,
electromagnetic waves etc., the function $G(u,v)$ satisfies the wave equation
$G(u,v)_{uv} =0$, when working in the Einstein frame. For general matter fields, however,
 this isn't necessary the case. The $v$ and the
$u$  Penrose limits of the above spacetime are  easily found by taking $u=0$ and $v=0$,
or, again, by dropping the dependense of the metric on $u$ or on $v$.
\footnote {To be more formal, one can always introduce scaling parameters $a$ and $b$ into the
co-ordinates $u$ and $v$ and then 
consider the limit as $a\to0$. 
I am grateful to Jerry Griffiths for prodding me to clarify this point.}
To write the plane wave in Rosen \cite{rosen} co-ordinates, after the limit is taken, one must rescale
the $u$ or $v$ co-ordinate with $f(u)du \to du$ or $f(v)dv \to dv$. The metric then becomes, say
in $u$-limit:

\be
ds^2 =- dudv + \, G(u) \, e^{p(u)}\, dx^2 \, + \, G(u) \, e^{-p(u)} \, dy^2
\ee 

The matter fields are treated by the same token. A similar scheme one follows
extending the Szekeres line-element into the plane wave regions. However, 
the evaluation of the Penrose limit is much  simpler procedure. In the  case of the collision of plane
waves, one must extend smoothly the interaction 
region into the plane wave regions  imposing  boundary conditions on the null hypersurfaces,
both for the gravitational and the matter waves.  In the case of the Penrose limit, one does
not extend the spacetime across the null hypersurfaces staying ``inside" 
the background spacetime, therefore there are no smoothness restrictions. 

A different observation, because of the similarity of the two problems,
may be made about the uniqueness relation between the interaction region, and the two 
$u$ and $v$ null limits. Given both  $u$ and $v$-limits, and the first two derivatives
of the metric in these limits, and the field equations, they {\em uniquely} determine 
the whole spacetime, 
this is because the initial value problem in the context of the colliding wave problem is
 well posed and defined: given the data on the null boundaries (Penrose limits, and the derivatives), the solution 
in the ``inside" 
region is unique. The above statement is of course true as long as the governing field equations are the 
low energy equations of the string theory, i.e. the Einstein Equations. In general, it is possible
that one would have to rely on higher derivatives when imposing the vanishing of the $\beta$ functions.

 Moreover, since Penrose limit is taken both in the metric and matter field 
components, and not on the level of phenomenological stress-energy tensors  as
 sometimes  done in the colliding wave problem, the {\em ambiguity problem} \cite{fein-mac-sen} of
possible non-unique evolution when the matter fields are not specified, but rather represented by
the algebraic form of the stress-tensor,
does not apply. The uniqueness of the Szekeres background, given the two Penrose limits, in the 
sense that the spacetime is completely determined by the two
 limiting metrics and the associated matter fields, may have important consequences for the string theory.
Till now, we  have had a rough relation between the whole space and a single plane-wave limit. Now, we have a 
one-to-one correspondence. Thus $AdS \times S$ background is completely determined by the two
``orthogonal" Penrose limits. Having only one limit does not determine the whole spacetime.
Thus, the two orthogonal Penrose limits form a sort of classical  {\em holographic boundary} for the
background with $N-2$ commuting Killing directions. 
To push this idea somewhat further we will give below an explicit example of a deformed spacetime
with one plane wave limit exactly as in  $AdS \times S$, but the other quite different.

 Exploiting the analogy between the problems, we can also comment about the hereditary 
properties in the sense of isometries. The situation is that generically plane waves have at
 least
a five-dimensional group of isometries in four dimensions. Upon collision this symmetry  is broken and
reduced to at least two Killing fields, in $x$ and $y$ directions in our notation. Vice versa, 
the two Killing directions are always trivially conserved in the null limit. Other commuting Killing
directions are also conserved.
Moreover, generically to the two Killing directions one must add another three in
the plane wave limit. For more elaborated discussion on this issue see ref. \cite{blau}.

As far as curvature singularity is concerned, generically the Szekeres geometries  have
curvature singularity at $G=0$, yet there is a class of solutions which instead of the  singularity 
at $G=0$ present a  Cauchy Horizon with possible non-Hausdorff extensions. We will comment about
these solutions  towards the end of the paper.

With all this in mind let us consider several examples which may be of interest in the string
theory setting. Our first example will be the spacetime which has Penrose v-limit exactly as the
$AdS_{2}\times S^{2}$, but its u-limit is Minkowski. This background, or target space, in the language
of string theory, may be thought of as a non-conformal deformation of $AdS_{2}\times S^{2}$,
or more precisely, 
of its Penrose limit. 
Surprisingly, however, the two different Penrose limits are exactly solvable string backgrounds,
one the Minkowski spacetime and the other the electromagnetic plane wave with the constant
profile, the one, one obtains in the $AdS \times S$ case.

 We will write the spacetime directly in Szekeres co-ordinates:

\be
f=\frac{\cos{bv}\sqrt{1-a^2\,u^2}}{\sqrt{\cos^{2}bv-a^2\,u^2}},\hspace{.2in} G=\cos^{2}bv-a^2\,u^2, \hspace{.2in}
p=\log{\frac{(1-au)}{(1+au)}}
\ee

with the electromagnetic field given (in the notation of \cite{griffiths}):

\be
\Phi_{0}=B \, \frac{b\cos{bv}}{\sqrt{\cos^{2}bv-a^2\,u^2}}, \hspace{.3in} \Phi_{2}=A \, \frac{a\sin{bv}}{(1-a^2u^2)\sqrt{\cos^{2}bv-a^2\,u^2}}
\ee

Here  $a$ and $b$ are constants and the functions $A$ and $B$ are defined via
 $l_{\mu}=A^{-1}\delta _{\mu}^{0}$, 
$n_{\mu}=B^{-1}\delta _{\mu}^{1}$ and the vectors $l_{\mu}$ and $n_{\mu}$ are null vectors
orthogonal to the planes spanned by the Killing vectors.

This solution was first integrated in \cite{griffiths2} and represents in the context of 
the colliding wave problem a region due to the interaction of a plane gravitational
 and plane electromagnetic wave. The parameter $a$ may be considered conformality parameter, as
$a\to 0$ we recover the conformal background.
 We now take the Penrose limits. The u-limit ($B_{v=0} \to 0$) is:

\be
ds^2=-2dudv+(1-au)^2\,dx^2+(1+au)^2\,dy^2, \label{grifleft}
\ee

 with no electromagnetic field present, while the v-limit ($A_{u=0} \to 0$) becomes:

\be
ds^2=-2dudv+\cos^{2}(bv)\,(dx^2+dy^2), \hspace{.2in} \Phi_{0}=b \label{grifright}
\ee

Both line elements are written in the so-called Rosen \cite{rosen} co-ordinates,
 where the metric is cast
in the following form:

\be
ds^2= -2dudv+d^2\,dx^2 + c^2\, dy^2,
\ee
here $d$ and $c$ are functions of $u$ or $v$ alone. To put the metric into the so-called 
Brinkmann-Peres form  \cite{brinkman}, \cite{peres}: 

\be
ds^2=-2dudr-(h_{11}X^2+h_{22}Y^2)\,du^2 + dX^2 + dY^2,
\ee
one must perform the following co-ordinate transformation

\be
X=d\,x, \hspace{.2in} Y=d\,y,\hspace{.2in} r=v+\half dd' \, x^2 + \half cc' \,y^2,
\ee
here $'$ denotes derivative with respect to the argument, and $d$, $c$, $h_{11}$ and $h_{22}$ are
related by:

\be
d''=-h_{11} \, d,\hspace{.2in} c''=-h_{22} \, c.
\ee

We note therefore, that the Brinkmann-Peres wave amplitudes $h_{11}$ and $h_{22}$ can be 
directly read off from the Rosen form. Incidentally, the Weyl and the
energy momentum tensor components can be expressed directly via the Rosen functions $d$ and
$c$ as \cite{griffiths}

\be
\Psi_{4}=-\half \frac{cd''-dc''}{dc},\hspace{.3in}\Phi_{22}=-\half \frac{cd''+dc''}{dc}.
\ee

Consequently, returning to our solution,
  the limit (\ref{grifleft}) is just  Minkowski spacetime, while the limit (\ref{grifright})
gives:

\be
ds^2= -2dudr-b^2( X^2+Y^2)du^2 + dX^2 + dY^2 \label{c-w}
\ee

Both limiting spacetimes play central role in  recent investigations of the string theory.
The Minkowski background for obvious reasons, but the  line element (\ref{c-w})  has a very special
relation to  string theory. It belongs to a class of the so-called Cahen-Wallach spacetimes
\cite{cahen-wallach} with constant negative eigenvalues ${-b^2}$. The higher dimensional analogs
of these spacetimes were studied as exactly solvable string theories in Ramond-Ramond background.
We therefore conclude here, that if there is any deep relation between the plane-wave limits
and the underlying target space  geometry, this must be inferred from the two orthogonal Penrose limits, rather
than from a single one. The existence of curvature singularity in a deformed target space signals
yet another problem. We will comment about this below.

Let us consider now the Penrose limits of the so-called Nappi-Witten cosmological
solution. 
The four-dimensional cosmology studied by Nappi and Witten \cite{n-w} results 
as the target space theory of a $SL(2,I\!\!R)\times SU(2)/I\!\!R\times U(1)$ gauged 
Wess-Zumino-Witten model. The solution contains, besides the metric, non-trivial values
for the dilaton and antisymmetric tensor field. The solution containing the 
non-vanishing B-field can be obtained by an 
$O(2,2;I\!\!R)$ rotation of the metric given below \cite{g-p,g-m-v} (for a review see \cite{g-p-r}).

The  typical Nappi-Witten line element belongs to the general class of closed
inhomogeneous dilaton cosmologies \cite{cchfein} and  is given by:

\begin{eqnarray}
ds^{2}=-dt^{2}+dw^{2}+\tan^{2}{w}\, dx^{2}
+\cot^{2}{t} \, dy^{2},
\label{gpr}
\end{eqnarray}
together with the dilaton field
\begin{eqnarray}
\phi=\phi_{0}-\log (\sin^{2}{t}\,\cos^{2}{w})
\label{dill}
\end{eqnarray}

The above line element is written in string frame and may be thought of as a  product of two two-dimensional 
black holes with Euclidean and Lorentzian signatures, both being 
 exact string backgrounds \cite{witten}, corresponding  to a
$SL(2,I\!\!R)/SO(1,1))\times SU(2)/U(1)$ coset model. Due to the presence of the non-trivial
dilaton field the Einstein and the string frames are not exactly equivalent and to study
the Penrose limit it is convenient here to stick to the string frame. The line element is already
written in Szekeres co-ordinates (just take $t+w=u$ and $t-w=v$) and it is easy to see
 that both $u$ and $v$ limits are
equivalent. Taking say the $u$- limit $v \to 0$, we obtain:

\be
ds^2=-2dudv+\tan^{2}{u\over 2} \, dx^2 +\cot^{2}{u\over 2} \, dy^{2}, 
\hspace{.3in} \phi=\phi'_{0}-2\log (\sin{u})
\ee

It is easy to see that the Penrose limit of the Nappi-Witten universe is the 
plane wave solution that may be obtained also from the gauged 
$\left[ SU(2) \times SL(2,I\!\!R)  \right]/ \left[U(1)  \times I\!\!R)\right]$
WZW model as given in \cite{sfetsos}. 
 
The harmonic forms of the wave amplitudes
 $h_{11}$ and $h_{22}$ are
given by:

\be
h_{11}=-2/cos^{2}{u\over 2},  \hspace{.3in} h_{22}=-2/sin^{2}{u\over 2}
\ee

It is interesting to see how the Penrose limits of this model differ from the one 
evaluated for the FRW cosmology. Let us consider the spatially flat model for simplicity.
The models with non-zero spatial curvature may be written in Szekeres-like co-ordinates
along the lines of the work \cite{feinvazq}, where one may also find the higher dimensional non-singular
generalisations of these models.
The line element for the spatially flat 
isotropic  dilaton solution of
 Einstein field equations  is given  by:

\be
ds^2= t \, (-dt^2 + dx^2 +dy^2 +  dz^2) \hspace{.3in} \phi=\frac{\sqrt{3}}{2}\log{t}
\ee 

The string frame metric is obtained by globally multiplying the line element by $e^{2\phi}$.
Introducing a pair of null co-ordinates as before, taking the $v=0$-Penrose limit
and rescaling the co-ordinate $u$, we get:

\be
ds^2= -2dudv \, + \, u^{\frac{1}{\sqrt{3}+2}} \, (dx^2 +dy^2 )
\ee

The harmonic functions $h_{11}$ and $h_{22}$  are then
given by:

\be
h_{11}=h_{22}=\frac{3+2\sqrt{3}}{4 \, (2+\sqrt{3})^{2}} \, u^{-2},
\ee

in agreement with \cite{blau}. It is not clear as to whether this limiting behaviour is
of some  interest in the stringy context.

As mentioned above, the Szekeres geometry is singular at $G=0$.  However, there exist a class of 
solutions \cite{feinib} where instead of curvature singularity at $G=0$ there appears the so-called
Cauchy Horizon.  The manifold is therefore non-Hausdorff, in the sense that there exist 
different non-equivalent analytic extensions across the Horizon.
Near the curvature singularity the geometry behaves
as a Kasner model (see for example \cite{fkvm} and \cite{bv} where these models were studied in the 
context of the pre big bang cosmology).
Near the Cauchy Horizon, on the other hand, the line element has degenerate Kasner exponents 
$(0,1,0)$:

\be
ds^2 \sim -2 \, dudv \, + \, (a-u-v)^2 \, dx^2 \, + dy^2, \label{horizon}
\ee

which we recognize as Milne universe ($t=a-u-v$ and $z=u-v$ etc.)  The models with Cauchy Horizon
were not considered of interest in the context of the pre big bang cosmology \cite{veneziano} due to the
belief that one must have  strong coupling regime near the singularity. Another crucial point was that
the singular models were related to the pbb inflation  to resolve the usual cosmological problems.
 In the re-newed attempt
to have a ``go" at the pbb idea \cite{khoury}, 
it is exactly the weak coupling behaviour near the $t \to 0$, 
corresponding to regular dilaton leading
  to the Killing Cauchy Horizon, that is preferred. 
It  remains to be seen how
these models may tackle the real cosmological problems (see though \cite{linde}), 
yet the singularity problem in these models is of a different kind and there are some new
ideas as to how to approach the resolution of this problem \cite{seiberg} in string theory.

To construct backgrounds with such properties (Cauchy Horizons),
consider the following dilaton solution of Einstein equations:

\be
ds^2=- f(t,\theta) \, (dt^2 - d\theta^2 ) + G(t,\theta)\,(e^{p(t,\theta)}\,d\varphi^2 +e^{-p(t,\theta)}\,d\psi^2),
\ee

with

\be 
f= (\cos^2 {t}-\cos^2 {\theta}) \, e^{b^2\, G^2}  , \qquad G=\, \sin{\theta}\sin{t}, \qquad p= \, \log{G}, 
\ee
along with

\be
\phi=b \, \cos(\theta)\cos(t) ,
\ee
and the co-ordinates are taken to be Euler angles. An infinite dimensional families
of such ``non-singular" solutions may be constructed  by for example choosing the scalar
field in terms of regular Legendre functions $P_{l}(\cos{t})\,P_{l}(\cos{\theta})$, or their 
linear combinations,
and
then integrating the function $f$, keeping $p$ and $G$ unchanged along the lines of \cite{cchfein}
(see as well \cite{feinvazq} in M-theory context).
Note that without the dilaton ($b \to 0$), the solution is
 flat. The dilaton is regular at $G=0$. The regularity
of the dilaton introduces an essentially different behaviour as compared to the Nappi-Witten background.
The near Horizon geometry  ($\phi \to b$, $f \to \cos^2 {t} \,- \, \cos^2 {\theta}$) is given by
(\ref{horizon}) and is Milne, and both orthogonal Penrose limits near the Horizon are  of course flat.

The  telling point, apart from being useful for modeling different approaches to
resolution of the singularity problem,
is that these non-Hausdorff 
time dependent backgrounds could be used to understand the definition of
quantum vacuum states, one of the central
problems of the string theory on non-trivial time dependent backgrounds.
The situation resembles that of quantizing matter
fields on the spacetimes with Killing-Cauchy Horizons \cite{verdaguer}, \cite{fein-ps}. 
What one does in these situations, is to define the ``in" vacuum on the Penrose boundaries. The ``out"
vacuum can be 
defined with the help of the null Killing fields of the  Killing Cauchy Horizon. Note, that
the  existence of the preferred vacuum state, invariant under the symmetries associated
with the Killing-Cauchy Horizon, follows from the Kay and Wald theorem \cite{kay}, who also have
shown that this preferred state is {\em unique}. We believe that this is a path worth being explored
in details in the future.

To conclude, we have shown that there is an intimate relation between the concept of
Penrose limit and the problem of classical relativity which studies the outcome of the collision
between the plane waves. This relation was exploited to shed some light on several issues of 
the string theory. We have argued that with certain symmetries given, the spacetime, the arena
for string propagation, is uniquely determined by the two orthogonal Penrose limits along with a couple of
derivatives of the metric in these limits. This of course is true as long as the pertinent equations of the
theory are the Einstein equations, i.e. the low energy equations of the string theory. In the full 
fledged string theory, one would probably need to rely on higher derivatives as well. It would be interesting
to consider this question with more rigor in future.
We have shown the
existence of non-Housdorff backgrounds with Killing Cauchy Horizons instead
of curvature singularities, symmetries of which can be
used to determine the vacuum state of the string theory. We have given a simple way to evaluate
Penrose limits, and have done so for several examples relevant to string theory.

\centerline{\bf Acknowledgments}
I am grateful to J.B. Griffiths and J.L. Ma\~nes for valuable comments and suggestions.
This work was supported by the University of the Basque Country Grants UPV 172. 310-GO 2/99 and
The Spanish Science Ministry Grant 1/CI-CYT 00172. 310-0018-12205/2000.

\end{document}